\documentclass[aps,prd,nofootinbib,longbibliography,reprint]{revtex4-1}
\usepackage{latexsym,bm,graphicx,color,xcolor,nicefrac,titletoc,dirtytalk,enumerate,mathtools,xfrac,xcolor,tensind,physics}

\usepackage{newtxtext,newtxmath,newtxtt}

\DeclareMathAlphabet{\mathcal}{OMS}{cmsy}{m}{n}

\usepackage{hyperref}
\hypersetup{linktoc=all,colorlinks=true,citecolor=blue,linkcolor=blue,filecolor=blue,urlcolor=blue}

\renewcommand{\thesection}{\arabic{section}}
\renewcommand{\thesubsection}{\arabic{section}.\arabic{subsection}}
\renewcommand{\thesubsubsection}{\arabic{section}.\arabic{subsection}.\arabic{subsubsection}}

\usepackage{titlesec}
\titleformat{\section}
{\centering\normalfont\bfseries\sffamily}{\thesection}{0.3em}{}
\titleformat{\subsection}
{\centering\normalfont\normalsize\bfseries}{\thesubsection}{0.3em}{}
\titleformat{\subsubsection}
{\centering\normalfont\normalsize\bfseries}{\thesubsubsection}{0.3em}{}

\numberwithin{equation}{section}

\def\dif{\mathrm{d}}

\newcommand{\p}{\partial}
\newcommand{\m}{\mu}
\newcommand{\n}{\nu}
\newcommand{\nab}{\nabla}
\def\d{\delta}
\def\D{\Delta}

\def\L{\Lambda}

\newcommand{\f}{\phi}

\newcommand{\vf}{\varphi}

\newcommand{\cL}{\mathcal{L}}

\newcommand{\cF}{\mathcal{F}}

\newcommand{\bR}{\bar{R}}

\newcommand{\bg}{\bar{g}}
\newcommand{\bnab}{\bar{\nabla}}
\newcommand{\tT}{\widetilde{T}}

\begin{document} 
\title{\sffamily Regularized Weyl double copy}

\author{G\"{o}khan Alka\c{c}}
\email{alkac@mail.com}
\affiliation{Department of Software Engineering, Faculty of Engineering and Architecture,\\ Ankara Science University, 06200 Ankara, Turkey}

\author{Mehmet Kemal G\"{u}m\"{u}\c{s}}
\email{kemal.gumus@metu.edu.tr}
\affiliation{Department of Physics, Faculty of Science and Letters,\\Middle East Technical University, Çankaya 06800, Ankara, Turkey}

\author{Oğuzhan Ka\c{s}\i k\c{c}\i }
\email{kasikcio@itu.edu.tr}
\affiliation{Department of Physics Engineering, Faculty of Science and Letters,\\ Istanbul Technical University, Maslak 34469 Istanbul, Turkey}

\author{Mehmet Ali Olpak}
\email{maolpak@thk.edu.tr}
\affiliation{Department of Electrical and Electronics Engineering, Faculty of Engineering,\\ University of Turkish Aeronautical Association, 06790, Ankara, Turkey}

\author{Mustafa Tek}
\email{mustafa.tek@medeniyet.edu.tr}
\affiliation{Department of Physics Engineering, Faculty of Engineering and Natural Sciences,\\ Istanbul Medeniyet University, 34000, Istanbul, Turkey}

\begin{abstract}
We propose a regularization procedure in the sourced Weyl double copy, a spinorial version of the classical double copy, such that it matches much more general results in the Kerr-Schild version. In the regularized Weyl double copy, the anti-de Sitter (AdS) and the Lifshitz black holes, which form the basis of the study of strongly coupled gauge theories at finite temperature through the AdS/CFT correspondence and its non-relativistic generalization, become treatable. We believe that this might pave the way for finding out a relation between the classical double copy and holography.
\end{abstract}
\maketitle
\section{Introduction}
The classical double copy is a map between certain exact solutions of general relativity (GR) and Maxwell's theory of electrodynamics whose mere existence should surprise anyone at first sight considering the nonlinear nature of GR. A more natural relation could be expected with Yang-Mills (YM) theory, a non-linear generalization of Maxwell's theory with interacting photons, which is, indeed, the historical path followed to obtain the double copy relations.

Just like many other important developments in high energy physics, the first hint to the double copy arose from string theory when Kawai, Lewellen and Tye (KLT) showed that any closed string amplitude at the tree level can be expressed as a linear sum of products of two open string amplitudes \cite{Kawai:1985xq}. Since, according to string theory, gravitational and gauge interactions are described by closed and open strings respectively, the KLT result suggests that, in the field theory limit, one should obtain a set of relations between the scattering amplitudes of Yang-Mills theory and the massless sector of the closed strings, i.e., the graviton, a dilaton and a two-form field. Such a version of the double copy was discovered by Bern, Carrasco and Johansson \cite{Bern:2008qj}, which proved to be very useful in computing quantum gravity amplitudes at the tree-level and also allows the extension to the loop level \cite{Bern:2010yg,Bern:2010ue,Bern:2010ue,Bern:2010tq,Carrasco:2011mn,Oxburgh:2012zr,Bern:2013yya}. In addition to the realization that perturbative classical spacetimes can be obtained by squaring the numerator of some diagrams in YM theory \cite{Luna:2016hge}, the classical double copy as a map between exact solutions of GR and Maxwell's theory was discovered \cite{Monteiro:2014cda,Luna:2018dpt} and various aspects of it has been extensively studied in the literature (see \cite{Bern:2019prr,Kosower:2022yvp,Adamo:2022dcm} for extensive reviews).

Among the two exact versions of the classical double copy, the Kerr-Schild (KS) double copy (KSDC) is based on the discovery of G\"{u}rses and G\"{u}rsey \cite{Gurses:1975vu} and its generalizations \cite{Taub:1981evj,Xant1,Xant2}  that the Ricci tensor with mixed indices becomes linear in the perturbation when the spacetime metric is written in the KS coordinates around a background spacetime (see \cite{Stephani:2003tm} for a summary of important properties). In this paper, we will consider the most general formulation of the KSDC presented in \cite{Alkac:2021bav}. For a metric written in the KS form
\begin{equation}
	g_{\m\n}=\bar{g}_{\m\n}+\phi\, k_\m k_\n,\label{KS}
\end{equation}
where $\phi$ is a scalar and the vector $k_\m$ is null and geodesic with respect to both the full metric $g_{\m\n}$ and the background metric $\bar{g}_{\m\n}$, one can study the trace-reversed  Einstein's equations in $d$ dimensions with a cosmological constant
\begin{equation}
	R^\m_{\  \n} -\frac{2\,\Lambda}{d-2}\, \delta^{\m}_{\  \n}=\tT^\m_{\  \n}=T^\m_{\  \n} -\frac{1}{d-2}\,\delta^\m_{\  \n}\,T,\label{reversed}
\end{equation}
and find the following scalar-gauge field equations defined on the background spacetime\footnote{There are some additional terms in these equations that vanish for all the examples that have been studied so far. We do not include them here for simplicity.}
\begin{equation}
	\bar{\nabla}_{\n} F^{\n \mu} =J^\m, \qquad \bar{\nabla}^2 \phi + \mathcal{C}  = j,\label{scalargauge}
\end{equation}
where one makes use of a Killing vector of both $g_{\m\n}$ and $\bar{g}_{\m\n}$, i.e.,
$\nab_{(\m}V_{\n)} = \bnab_{(\m}V_{\n)} = 0$. The gauge theory field strength tensor $F_{\m\n}= 2 \p_{[\m} A_{\n]}$ is formed from the gauge field $A_{\m} = \phi\, k_\m$, and for each solution of Einstein's equations \eqref{reversed} that can be written in the KS coordinates as in \eqref{KS}, one can construct a solution to scalar-gauge field equations \eqref{scalargauge} defined on the background spacetime with the metric  $\bar{g}_{\m\n}$. The $\mathcal{C}$ term in \eqref{scalargauge} is the usual modification to Poisson's equation in curved spacetime (see \cite{Alkac:2021bav} for details). While the gauge field $A_\m$ is called the single copy, the scalar $\phi$ is named as the zeroth copy.

The gauge theory source gets contributions from both gravitational sources described by $\tT^\m_{\  \n}$ and from the so-called deviation tensor $\D^\m_{\  \n} =\bR^\m_{\  \n}-\frac{2\,\L}{d-2}\, \d^\m_{\ \n},$ which characterizes the deviation of the background spacetime from a constant curvature spacetime satisfying the Einstein equations in vacuum. In this paper, we will focus on static black hole solutions for which the time-like Killing vector $V^\m = \d^{\m}_{\ 0}$ is used. For such a case, the sources are given by
\begin{equation}
	J^\m =  2 \left[\D^\m-\tT^\m\right],\qquad j=J_0=\bg_{0 \m} J^\m,\label{Jdef}
\end{equation} 
where $\D^\m = \D^\m_{\ 0}$ and $\tT^\m = \tT^\m_{\ 0}$. The $\D^\m$ term arising from the deviation tensor provides a geometric explanation to the constant charge density that fills all space and creates a radial, linearly increasing electric field, which is observed when an asymptotically AdS solution is studied on Minkowski background, and, in general, exhibits quite non-trivial features when the background spacetime is not maximally symmetric.

The second version, the Weyl double copy (WDC) \cite{Luna:2018dpt, Godazgar:2020zbv}, relies on some previous results on Petrov type D and N spacetimes admitting a Killing spinor \cite{Walker:1970un, Hughston:1972qf, Dietz361}, according to which there exists a particular relation between the completely symmetric Weyl spinor $\Psi_{A B C D}$ corresponding to a type D or N vacuum solution of 4$d$ Einstein equations (with $\Lambda = 0$) and the symmetric field strength spinor $f_{A B}$ corresponding to a solution of Maxwell's equation defined on the curved spacetime characterized by the Weyl spinor $\Psi_{A B C D}$, where the time component describes a scalar field satisfying the Poisson equation on the same curved background spacetime. It turns out that one has exactly the same equations when the scalar and the gauge fields are defined on the flat background spacetime when the curved spacetime metric can be written in the KS coordinates around the flat background. As a result, one obtains a second, spinorial version of the double copy. It is possible to derive this version of the double copy from twistor theory \cite{White:2020sfn} and develop a deeper understanding. For example, when the momentum space origins of the double copy are considered, it is already quite surprising to have a local relation in position space. Using twistor techniques for type D spacetimes, this was shown to be a consequence of the very special properties in the momentum space, which could be possible only for algebraically special spacetimes \cite{Luna:2022dxo}.

In a recent letter \cite{Easson:2021asd}, it was shown that, although not covered in the original theorems that inspired the construction of the WDC, sources on the gravity side might be handled term-by-term by considering a sum of scalar-gauge theories and the sourced WDC (SWDC) takes the following form
\begin{equation}
	\Psi_{A B C D} =  \sum_{i=1}^N \frac{1}{S_{(i)}}f^{(i)}_{(A B}f^{(i)}_{C D)}.\label{SWDC}
\end{equation}
Here, the $i=1$ term represents the scalar-gauge theories corresponding to the vacuum solution $\bar{\nabla}_{\n} F^{\n \mu}_{(i=1)} =0$ and other terms satisfy a sourced Maxwell equation $\bar{\nabla}_{\n} F^{\n \mu}_{(i>1)} =J^\m_{(i>1)}$, i.e., for each term in the metric other than the vacuum part, one finds a sourced Maxwell equation with a source proportional to the one described by the KSDC procedure. The complex scalar $S$  obeys the Poisson equation and, in general, the zeroth copy $\phi$ is a linear combination of its real and imaginary parts. So far, this has been applied to the Kerr-Newman black hole solution, the charged C-metric, and the most general type D solution of Einstein-Maxwell theory \cite{Easson:2022zoh}. This proposal can be checked in a non-trivial way by solving gauge field equations term-by-term according to the KS prescription and it was also shown that one loses the connection to the KSDC if a single product of spinor fields are used. 

For type D spacetimes, in a suitable spinor basis $\left\{o_A, \iota_B\right\}$, the Weyl spinor is given by $\Psi_{A B C D} = 6 o_{(A} o_B \iota_C \iota_{D)} \sum_{i}  \Psi^{(i)}_{2}$, where $\Psi^{(i)}_{2}$ is the only non-zero Weyl scalar computed for the part of the metric corresponding to the $i$-th source (we follow the conventions summarized in \cite{Easson:2022zoh}). In this paper, we will consider only static black hole solutions, for which many properties of the single copy solution take a simple form. The single copy field strength spinor reads $f_{A B} = Z\, o_{(A} \iota_{B)}$ where $Z$ is real and directly related to the radial single copy electric field $e=F_{rt}$. The generally complex field $S$ is also real and, therefore, can be identified with the zeroth copy $\phi$. As a result, in this relatively simple setup, it suffices to check the following consistency condition
\begin{equation}
	\Psi^{(i)}_{2} \propto \frac{Z_{(i)}^2}{\phi_{(i)}},\label{CC}
\end{equation}
term-by-term for the consistency of the proposal of \cite{Easson:2021asd} for the SWDC with the KSDC.

When this procedure is applied to a metric of the following form with a flat background metric 
\begin{align}
	\dd{s}^2 &= -h(r) \dd{t}^2 + \frac{\dd{r}^2}{h(r)}+r^2\dd{\Omega}^2\label{met}\\
	h(r) &= 1 + \sum_{n} \frac{a_n}{r^n}\label{h}\\
	\dd{\bar{s}}^2 &= - \dd{t}^2 + \dd{r}^2+r^2 \dd{\Omega}^2\label{back},
\end{align}
where $\dd{\Omega}^2 = \dd{\theta}^2 + \sin^2 \theta \dd{\phi}^2$ as first done in \cite{Easson:2021asd}, one finds
\begin{align}
	\phi &= -\sum_{n} \frac{a_n}{r^n}, \qquad e= \sum_{n} \frac{n a_n}{r^{n+1}} \propto Z\\
	\Psi_2 &= \sum_{n} \frac{ (n+1)(n+2) a_n}{12\,r^{n+2}}\label{psi2}.
\end{align} 
We see that the consistency condition \eqref{CC} is satisfied for asymptotically flat solutions ($n>0$). For the Reissner-Nordstr\"{o}m (RN) black hole solution of Einstein-Maxwell theory, one has $a_1 = -2M$, $a_2 = \frac{Q^2}{4}$, where $M$ and $Q$ are the black hole's mass and electric charge, respectively, and the scalar-gauge theory equations are satisfied term-by-term exactly as described by the SWDC in \eqref{SWDC}.
\section{Problems}
We are now in a position to discuss some problems regarding the proposal for the SWDC. We would like to emphasize that we do not claim that there is a pathology in the procedure as given in \cite{Easson:2021asd}; however, there are issues that need to be resolved for matching the SWDC with more general results in the KS side of the double copy.

First of all, since the Weyl tensor $W_{\m\n\rho\sigma}$ transforms homogeneously under a conformal transformation of the metric ($\tilde{g}_{\m\n} = e^{\psi} g_{\m\n}, \tilde{W}_{\m\n\rho\sigma} = e^{\psi}W_{\m\n\rho\sigma}$), the Weyl tensor and the Weyl spinor vanish for conformally flat spacetimes. Therefore, when one applies the same prescription for the SWDC to a term in the metric function which is conformally flat when considered alone, the procedure seems to break down since the corresponding contribution to the Weyl spinor vanishes and obviously cannot match a a non-zero relevant contribution in the KSDC. Such a phenomenon cannot be observed when working with asymptotically flat spacetimes since it corresponds to $n=-1, -2$ in \eqref{h} for which the contribution to the Weyl scalar $\Psi_2$ vanishes as can be seen in \eqref{psi2}. Hence, the resolution of this issue is crucial for the correct formulation of SWDC for type D solutions that are not asymptotically flat.

A related problem appears for the RN-AdS$_4$ black hole solution of Einstein-Maxwell theory with a negative cosmological constant described by the Lagrangian $\cL = \sum_{i} \cL_{i}$ with $\cL_{(i=1)} = R$, $\cL_{(i=2)} = - \frac{1}{4} \cF_{\m\n} \cF^{\m\n}$, and $\cL_{(i=3)} = -2 \L$. The metric function in \eqref{h} reads
\begin{equation}
	h(r) = 1 - \frac{2M}{r} + \frac{Q^2}{4 r^2} - \frac{\L r^2}{3}.
\end{equation}
According to \cite{Easson:2022zoh}, the single copy solution should be defined in a ``suitable flat space limit", which is the AdS$_4$ spacetime in the global static coordinates with the line element
\begin{equation}
\dd{\bar{s}}^2= -\left[1-\frac{\L r^2}{3}\right]\dif t^2+\left[1-\frac{\L r^2}{3}\right]^{-1} \dif r^2+r^2 \dd{\Omega}^2.\label{ads4}
\end{equation}
This is a natural choice that we obtain when the mass $M$ and the charge $Q$ are set to zero.  Applying the machinery of the SWDC, we find
\begin{align}
	\f&=\frac{2 M}{r}-\frac{Q^2}{4 r^2}, \qquad e =-\frac{2 M}{r^2}+\frac{Q^2}{2 r^3} \propto Z,\label{singRNads}\\
	\Psi_2^{(i=1)} &= -\frac{M}{r^3}, \qquad  \Psi_2^{(i=2)} = \frac{Q^2}{r^4}, \qquad \Psi_2^{(i=3)} = 0,\label{weylRNads}
\end{align}
where the single copy properties \eqref{singRNads} were first given in \cite{Carrillo-Gonzalez:2017iyj}. We see that although the contribution from the conformally flat part of the metric that arises due to the cosmological constant vanishes, the consistency condition \eqref{CC} is still satisfied since the cosmological constant has no effect on the properties of the single copy solution \eqref{singRNads} defined on the AdS$_4$ spacetime.

On the other hand, the same solution can be put into the KS form \eqref{KS} with a flat background metric given in \eqref{back}. This time, the single copy properties get a contribution from the cosmological constant as follows \cite{Alkac:2021bav}
\begin{equation}
	\f=\frac{2 M}{r}-\frac{Q^2}{4 r^2}+\frac{\Lambda  r^2}{3},\qquad e = -\frac{2 M}{r^2}+\frac{Q^2}{2 r^3}+\frac{2 \Lambda  r}{3} \propto Z,
\end{equation}
while the Weyl spinor takes the same form characterized by the Weyl scalars given in \eqref{weylRNads}. The consistency condition \eqref{CC} is obviously not satisfied by the contribution to the metric function from the cosmological constant, which is just the $n=-2$ term in \eqref{h}. For a complete equivalence of the KSDC and the SWDC, one would expect to cover both single copies in the SWDC.

One final problem is that some solutions to matter coupled GR do not have a vacuum part, i.e., when the matter coupling is turned off there remains no solution that both carries the symmetries of the ansatz that is used to derive the solutions and satisfies the field equations at the same time. It is desirable to understand whether (and, if yes, how) the SWDC can be realized in such cases.
\section{Resolution by regularization}
For the resolution of the first two problems, we are inspired by a simple observation for the RN-AdS$_4$ black holes: Although the standard procedure fails when the background metric is taken to be flat, the radial dependence of the Weyl scalar is still correct [$\Psi_2^{(i=3)} \propto \text{constant}$ apart from the $(n+2)$ factor that produces zero] such that the consistency condition \eqref{CC} is satisfied with $\phi_{(i=3)} \propto r^2$ and $e_{(i=3)} \propto r$. Motivated by this, we propose a three-step regularization procedure for the SWDC such that it will still work when one has a conformally flat part in the metric function that has a non-trivial effect on the single copy solution: Let us say the conformally flat part is of the form $\frac{a_{n_*}}{r^{n_*}}$ such that for $n=n_*$, the contribution to the Weyl scalar $\Psi_2^*$ vanishes. To cure this, one should proceed as follows:
\begin{enumerate}
	\item Take the problematic term as $\frac{a_{n_*}}{r^{n}}$, i.e., use an arbitrary exponent $n$ instead of $n_*$.
	\item Let $a_{n_*} \to \frac{a_{n_*}}{n-n_*}$ and calculate $\Psi_2^*$ using this coefficient.
	\item Set $n = n_*$ at the end.
\end{enumerate}
This way, the $(n-n_*)$ term in the Weyl scalar $\Psi_2^*$ is removed in a controlled manner and a non-zero contribution is obtained. For a metric of the form \eqref{met}, the consistency condition \eqref{CC} is automatically satisfied and the problematic term in the RN-AdS$_4$ black hole with a flat background metric corresponds to $n_* = -2$. Although we are not aware of a solution of matter coupled GR with an $n_* = -1$ term, which is of the form \eqref{met}, such a problem would be easily solved.

When considered around a flat background metric,  the procedure for the RN-AdS$_4$ black hole works as follows: Taking the term with a cosmological constant in the metric function as $h_{(i=3)} = a_{-2}\, r^n$ with $a_{-2} = -\frac{\Lambda}{3}$, the Weyl scalar becomes $\Psi_2^{(i=3)}=\frac{(n+1) a_{-2}}{12 r^{n+2}}$. However, letting $a_{-2} \to \frac{a_{-2}}{n+2}$ and then setting $n=-2$ yields $\Psi_2^{(i=3)}=\frac{\Lambda}{36}$, which satisfies the consistency condition \eqref{CC} for $i=3$.

One might rightfully argue that this form of the metric ($g_{tt} g_{rr} = -1$) is too simple to check the validity of the regularized Weyl double copy (RWDC). Therefore, we put it to the test with a set of examples that requires the most general formulation of the KSDC summarized at the beginning, i.e.,  the Lifshitz black holes. As a by-product, they will show us how the third problem must be handled.
\section{Lifshitz black holes}
In order to study a scenario as general as possible, we will consider the following ansatz for Lifshitz black holes with different horizon topologies
\begin{equation}
	\dif s^2 = L^2 \left[ -r^{2z} h(r) \dif t^2 +\frac{\dif r^2}{r^2 h(r)} + r^2 \dd{\Sigma}_k^2\right],\label{lif}
\end{equation} 
where $k=+1, 0, -1$ correspond to 2$d$ spherical, planar and hyperbolic surfaces ($\text{S}^2$, $\text{E}^2$ and $\text{H}^2$) respectively, and $z$ is the dynamical exponent. When $z \neq 1$, one has an anisotropic scaling of the time coordinate in the boundary field theory and these solutions can be used to probe the properties of finite temperature non-relativistic systems holographically when $k=0$ \cite{Taylor:2015glc} ($z=1$ corresponds to the relativistic case). When written in the KS coordinates, the background spacetime is the Lifshitz spacetime whose line element can be found by setting $h(r) = 1$ in \eqref{lif}. For a generic metric function as in \eqref{h}, one has
\begin{align}
	\phi &= -L^2 \sum_{n} \frac{a_n}{r^{n-2z}}, \,\,\, e = L^2 \sum_{n} \frac{(n-2z) a_n}{r^{n-2z+1}} \propto r^{z-1} Z,\label{elif}\\
	 \Psi_2 &= - \frac{ k}{6 L^2 r^2} +\frac{z(z-1)}{6 L^2} + \sum_{n} \frac{(n-z)(n-2z+2)\, a_n }{12L^2r^n}.\label{psilif}
\end{align}
First of all, since the Lifshitz black holes cannot be obtained as vacuum solutions, all the non-zero $a_n$'s here can only be generated by matter coupling. Therefore, there is no analog of the mass term in the RN-AdS$_4$ black hole and no vacuum part of the metric. As a result; in the Weyl scalar $\Psi_2$ in \eqref{psilif}, there is a term that is independent of $a_n$'s and has no direct meaning regarding the properties of the single copy. Note that it vanishes for $z=1$ and $k=0$, which corresponds to the AdS$_4$ spacetime in Poincaré coordinates that is a vacuum solution. Other than this term, the consistency condition \eqref{CC} is satisfied with the data given above provided that a regularization is employed for the critical values $n_* = z$ and $n_* = 2z - 2$ while calculating the Weyl scalar $\Psi_2$. Note also that an additional difficulty might arise since one might now have zero electric field when $n=2z$ [see eq. \eqref{elif}] unlike for the class of metrics satisfying $g_{tt} g_{rr}=-1$. In such a case, a regularization with $n_* = 2z$, which is another critical value, should be used in the calculation of the electric field. In Table \ref{table}, we provide various examples.
\begin{widetext}
\begin{center}
\begin{table}[]
	\setlength{\tabcolsep}{20pt}
	\renewcommand{\arraystretch}{2}
	\centering
	\begin{tabular}{|c|c|c|c|}
		\hline
		I & $\cL_m = -\frac{1}{4} \mathcal{F}_{\mu \nu} \mathcal{F}^{\mu \nu} -\frac{1}{2} m^2 a_\m a^\m -\frac{1}{4} \mathcal{G}_{\mu \nu}\mathcal{G}^{\mu \nu}$ & $h(r)=1-\frac{q^2}{8 r^4}$ & $k=0$, $z=4$ \\
		\hline
		II & $\cL_m=\frac{1}{2} \partial_{\mu} \vf \partial^{\mu} \vf-\frac{1}{4} e^{\lambda \vf} \cF_{\mu \nu} \cF^{\mu \nu}$ & $h(r) = 1 - \left(\frac{r_+}{r}\right)^{z+2}$ & $k=0$, $z>1$ \\
		\hline
		III &   $\cL_m = -\frac{1}{4} \mathcal{F}_{\mu \nu} \mathcal{F}^{\mu \nu}-\frac{1}{12} \mathcal{H}_{\mu \nu \tau} \mathcal{H}^{\mu \nu \tau}-C \epsilon^{\mu \nu \alpha \beta} B_{\mu \nu} \mathcal{F}_{\alpha \beta}$ & $h(r) = 1 + \frac{k }{2 r^2}$ & $k\neq0$, $z=2$ \\
		\hline
	\end{tabular}
	\caption{Matter couplings, metric functions, $k$ and $z$ for three different Lifshitz black hole solutions considered in the text. $\varphi$ is a scalar field. $\cF_{\mu \nu} = 2 \partial_{[\m} a_{\n]}$ and $\ \mathcal{G}_{\m\n}$ are two-form fields. $\mathcal{H_{\m\n\tau}} = \partial_{[\mu} B_{\nu \tau]}$ is a three-form field. For examples I and II, the single copies were found in \cite{Alkac:2021bav}. Example III is studied for the first time here. Solutions are from \cite{Pang:2009pd,Taylor:2008tg,Mann:2009yx} respectively. The KS single copies for examples I and II were obtained in \cite{Alkac:2021bav}.}\label{table}
\end{table}	
\end{center}
\end{widetext}

In example I, one has a sourced single copy solution and the relevant quantities that are needed to check the consistency condition \eqref{CC} are as follows\footnote{There are some typos in \cite{Alkac:2021bav} but these properties follow from eqn.s (4.5) and (4.27) in this reference.}:
\begin{align}
	h_{(i=2)} &= - \frac{q^2}{8 r^4}, \qquad \phi_{(i=2)}= \frac{q^2L^2r^4}{8},\\
	e_{(i=2)} &= \frac{q^2L^2r^3}{2} \propto r^{3}Z_{(i=2)},\qquad \Psi_2^{(i=2)}=0,
\end{align}
The Weyl scalar $\Psi_2^{(i=2)}$ vanishes since $n=z=4$. However, this can be cured by a regularization with $n_* = z=4$. Taking $h_{(i=2)}=\frac{a_4}{r^n}$ with $a_4= -\frac{q^2}{8}$, one finds $\Psi_2^{(i=2)}=\frac{(n-4)(n-6)a_4}{12 L^2 r^n}$. We can now let $a_4 \to \frac{a_4}{n-4}$ and then set $n=4$, which yields
\begin{equation}
	\Psi_2^{(i=2)}=\frac{q^2}{48L^2r^4},
\end{equation}
which satisfies the consistency condition. 

Example II is the double copy of the vacuum solution of Maxwell's equations on the Lifshitz spacetime, which is the only known example in the KSDC where a non-vacuum gravity solution is mapped to a vacuum gauge theory solution in $d>3$ (3$d$ KSDC is a completely different story where this is a must \cite{CarrilloGonzalez:2019gof,Gumus:2020hbb,Alkac:2021seh,Alkac:2022tvc}). For this solution, which is valid for $z>1$, we have
\begin{align}
	h_{(i=2)} &= - \frac{r_+^{z+2}}{r^{z+2}}, \qquad \phi_{(i=2)}=L^2r_+^{z+2}r^{z-2},\\
	e_{(i=2)} &= (z-2)L^2r_+^{z+2}r^{z-3} \propto r^{z-1}Z_{(i=2)},\\ \Psi_2^{(i=2)} &=\frac{(z-4)r_+^{z+2}}{6L^2r^{z+2}}.
\end{align}
We see that the consistency condition \eqref{CC} is satisfied as long as $z\neq4$. Note that the $z=4$ case corresponds to the critical value $n_*=2z-2$ since $n=z+2$ in the relevant part of the metric function $h_{(i=2)}$. Taking $h_{(i=2)}= \frac{a_6}{r^n}$ with $a_6 = -r_+^6$, we have $ \Psi_2^{(i=2)} = \frac{(n-4)(n-6)}{12 L^2 r^n}$. Letting $a_6 \to \frac{a_6}{n-6}$ and then setting $n=6$ gives
\begin{equation}
	 \Psi_2^{(i=2)}=-\frac{r_+^6}{12 L^2 r^6},
\end{equation}
which now satisfies the consistency condition. 

The vanishing of the electric field for $z=2$ in this example can also be cured similarly. Taking $h_{(i=2)}=\frac{a_4}{r^n}$ with $a_4 = -r_+^4$ gives $e_{(i=2)}=\frac{(n-4) a_4}{r^{n-3}}$. Letting $a_4 \to \frac{a_4}{n-4}$ and then setting $n=4$ yields the electric field $e_{(i=2)}= - \frac{r_+^4}{r}$, which is compatible with the consistency condition.

Example III is a demonstration of the equivalence of two formulations of the classical double copy for a naked singularity where constant $t$ and $r$ surfaces are $\text{S}^2$, and a black hole with $\text{H}^2$ horizon. This is a particularly interesting case because the critical values $n_*=z$ and $n_*=2z-2$ coincide when $z=2$. The relevant quantities are
\begin{align}
	h_{(i=2)} &= \frac{k}{2 r^2}, \qquad \phi_{(i=2)}=-\frac{kL^2r^2}{2},\\
	e_{(i=2)} &= -k L^2 r \propto r \,Z_{(i=2)},\qquad \Psi_2^{(i=2)}=0.
\end{align}
In order the handle the vanishing of the Weyl scalar, we take $h_{(i=2)}=\frac{a_2}{r^n}$ with $a_2=\frac{k}{2}$, for which the Weyl scalar is given by $\Psi_2^{(i=2)}=\frac{(n-2)^2 a_2}{12 L^2 r^n}$. Since two of the critical values coincide, we have an $(n-2)^2$ factor, which can be cured by letting $a_2 \to \frac{a_2}{(n-2)^2}$ and then setting $n=2$. The Weyl scalar is obtained as
\begin{equation}
	\Psi_2^{(i=2)}=\frac{k}{24L^2r^2},
\end{equation}
satisfying the consistency condition \eqref{CC}. 

These examples cover all the critical values $n_*$ for which the SWDC does not seem to agree with the KSDC and we see that one can always ensure the agreement by the regularization procedure that we propose.
\section{Conclusions and outlook}
In this paper, by introducing a regularization procedure in the SWDC, we have shown how it can be made consistent with the KSDC when a certain part of the metric function in static black hole solutions is conformally flat but still affects the single copy properties nontrivially. Intriguingly, this turns out to be much more than curing a mathematical difficulty in the formulation. Handling this issue is directly related to the AdS and the Lifshitz black holes that form the basis of probing the properties of strongly coupled gauge theories at finite temperature holographically, \cite{Maldacena:1997re,Witten:1998qj,Gubser:1998bc, Taylor:2015glc} which is just another development that we owe to string theory. Since both the double copy and the AdS/CFT correspondence (together with its non-relativistic generalization) originate from string theory, we have a strong expectation that achieving consistency of different formulations of the classical double copy will be an important first step to understand whether the double copy ideas and holography are related in some way. While the gauge theory in the double copy lives in the same number of dimensions as the gravity theory, the gauge theory dual in holography is defined on the conformal boundary, which has one less dimension. Although one cannot point out a relation for now, the success of the WDC in capturing the asymptotic structure of spacetimes \cite{Godazgar:2021iae,Adamo:2021dfg,Mao:2023yle} together with our regularization procedure might be helpful in this regard.

We would like to note that, in all the examples considered here, the constant $a_{n_*}$ is proportional to either the cosmological constant $\Lambda$ or the Newton's constant $G$ which is hidden in the event horizon radius $r_+$. This causes one to think that our regularization procedure might have a field theoretical origin, which, we believe, deserves further study.

In our analysis, we have reached another important finding: When a solution does not have a vacuum part, then, on the left-hand-side of the SWDC formula \eqref{SWDC}, the leading term loses its meaning in the interpretation of the single copy. The Lifshitz black holes are the first kind of such examples but we expect the same behaviour in different types of solutions.

Despite the success of the RWDC that we have presented here, we would like to mention that the study of some solutions of $\mathcal{N}=0$ supergravity, the effective field theory emerging in the low energy limit of closed string theory, by using twistor methods has recently led to a double copy formula different than the one considered here \cite{Armstrong-Williams:2023ssz}. For a better understanding, more general and also different types of examples including rotating black holes and wave solutions should be studied.

\begin{acknowledgments}
	M. K. G. is supported by T\"{U}B\.{I}TAK Grant No 122F291.
\end{acknowledgments}
\bibliography{ref}
\end{document}